# Experimental observation of a photonic hook

Igor V. Minin[1,*], Oleg V. Minin[2], Gleb M. Katyba[3,4], Nikita V. Chernomyrdin[4,5,6], Vladimir N. Kurlov[3,5], Kirill I. Zaytsev[4,5,6], Liyand Yue[7], Zengbo Wang[7], and D. N. Christodoulides[8]

[1]Tomsk Politechnical University, Tomsk 634050, Russia
[2]Tomsk State University, Tomsk 634050, Russia
[3]Institute of Solid State Physics of RAS, Chernogolovka 142432, Russia
[4]Bauman Moscow State Technical University, Moscow 105005, Russia
[5]Sechenov First Moscow State Medical University, Moscow 119991, Russia
[6]Prokhorov General Physics Institute of RAS, Moscow 119991, Russia
[7]School of Electronic Engineering, Bangor University, Bangor, LL57 1UT, UK
[8]College of Optics/CREOL, University of Central Florida, Orlando, Florida 32816, USA
(Dated: July 5, 2018)

In this letter, we report the first experimental observation of a photonic hook – a new type of near field curved light generated at the output of a dielectric cuboid with a broken symmetry-having dimensions comparable to the electromagnetic wavelength. Given that the specific value of the wavelength is not critical once the mesoscale conditions for the particle are met, we here verify these predictions experimentally using a 0.25 THz continuous-wave source. The radius of curvature associated with the generated photonic hook is smaller than the wavelength used while its minimum beam-waist is about 0.44 $\lambda$. This represents the smallest radius of curvature ever recorded for any electromagnetic beam. The observed phenomenon is of potential interest in optics and photonics, particularly, in super-resolution microscopies, manipulation of particles and liquids, photolithography and material processing. Finally, it has a universal character and should be inherent to acoustic and surface waves, electron, neutron, proton and other beams interacting with asymmetric mesoscale obstacles.

The idea that light propagates along straight lines is known since antiquity. The development of Maxwell's electrodynamics further reinforced these notions by ensuring the conservation of electromagnetic momentum. The possibility that a wavepacket can freely accelerate even in the absence of an external force, was first discussed four decades ago, by Berry and Balazs [1] within the context of quantum mechanics. As indicated in [1], this is only possible as long as the quantum wavefunction follows an Airy-function profile. In 2007, this Airy self-acceleration process was first suggested and experimentally observed for the first time in optics by Siviloglou et al. [2, 3]. It is important to note that for finite power Airy beams, while the local intensity features do self-bend in a self-similar fashion, the Ehrenfest theorem still holds-thus preserving the balance of the transverse electromagnetic momentum. Ever since, this class of accelerating or self-bending beams has attracted considerable attention and found applications in many and diverse fields. In the past few years, other type of Airy-like accelerating curved beams have been intensely explored; among them: "half Bessel" [4], and Weber and Mathieu beams [5,6], to mention a few. In all cases, these Airy-like wavefronts propagate on a ballistic trajectory over a considerable distance while defying diffraction effects. They still remain the only example of a "curved light transport" in nature.

On the other hand, the interaction of light with transparent spherical particles has been heavily investigated over the years since the days of Pliny the Elder (AD 23–AD 79) [7]. By following the Mie formalism [8], one could easily analyze a transformation of the electromagnetic (EM) field structure from that produced by a small dipole (in the Rayleigh limit [9]) to a pronounced jet-like caustic formed near the outer part of a particle, as a function of its geometry, dimensions and refractive index [11]. In 2004, Chen et al. coined a new term "photonic nanojet" (PJ) for the sub-wavelength-scale focusing at the shadow side of a mesoscale dielectric particle [11]. By increasing the dimensions of a spherical particle the electromagnetic field structure evolves and tends to be more localized and directed forward [12]. Despite the beneficial performance of PJs in a number of applications, up to date, they have been generated primarily by symmetric dielectric particles, while no attention has been paid to asymmetric mesoscale obstacles or the formation of curved PJs.

In 2015, a new type of subwavelength curved PJ was proposed: a photonic hook (PH) [13]. This prospect was identified by means of numerical analysis of electromagnetic wave localization behind a dielectric three-dimensional (3D) particle with a broken symmetry [13,14]. But at present, to the best of our knowledge, no direct experimental confirmation of this phenomenon has ever been reported. This might be due to both difficulties in fabrication of micro-scale particles with high precision, as well as the limited resolution of existing 3D sub-wavelength-resolution imaging modalities in the visible range. Moreover, in a realistic optical band system, for handling the asymmetric dielectric particle, it would need to be in contact with a finite-size dielectric substrate, which generates its own scattered field, interacting with that of a particle, and, thus, leading to image distortions. Thereby, it could be a challenge today for experimental demonstration the PH phenomenon in the optical domain. However, taking into account that the specific value of the incident wavelength is not critical (as long as the mesoscale conditions of a particle are satisfied) [13], detailed emulation experiments with terahertz (THz) pave a way for advanced investigation, design and proof of concept demonstrations.

In this latter, we verify experimentally the existence of Minins' PH phenomenon for the first time. We demonstrate the PH effect using a continuous-wave scanning-probe microscopy, operating at 0.25 THz. Good agreement is found between our experimental results and our theoretical predictions.

For our demonstration, as a representative PH source, we use an asymmetric dielectric particle (ADP) with a simple geometry, which implies appending a triangular prism to the front side of a cuboid [13,14]; see Fig. 1 (a). We select a polymethylpentene (TPX, refractive index of $n$=1.46 at 0.25 THz) [15] as a material platform for fabrication of this particle. Priory to fabrication of the particle, we perform numerical analysis of the PH formation and optimized an angle of the appended prism $\theta$ and a size of the cuboid rib $L1$, using the commercial finite integral technique (FIT) software package – CST Microwave Studio© (CST), in order to achieve an expressive example of the PH behind the particle.

In Figs. 1 (c) and (d), we show an example of PJ formed behind the symmetric particle with an equal rib (Fig.1(b)) and examples of PH behind the asymmetric particles (Fig.1(c-e)). The phenomenon of focus bending the dielectric particle is caused by the interference of waves inside it as the phase velocity disperses. Due to the shape of particle with broken symmetry, the time of the complete phase of the wave oscillations varies irregularly in the particle [14]. As a result, the emitted light beam bends. It is worth to note that the PH is formed in the spatial region where the effects of near-field scattering (evanescent fields) play a significant role. The curvature of a photonic hook is defined by the factor $\alpha$ aided by a midline $L_{\text{jet}}$ (see Fig. 1 (a)), and is the angle between the two lines linking the start point with the inflection point, and the inflection point with the end point of PH, respectively, as defined in [14]. In Figs. 1 (c-e) we show results of the particle geometry optimization – i.e. an evolution of main PH parameters with changes in the particle geometry (top base length vs. bottom base length - $L1$ vs. $L2$). We found that the TPX cuboid with 4.8 and 6.4-mm-length top and bottom sides and $\theta$ = 18.25° should ensure generation of PH with the maximal curvature.

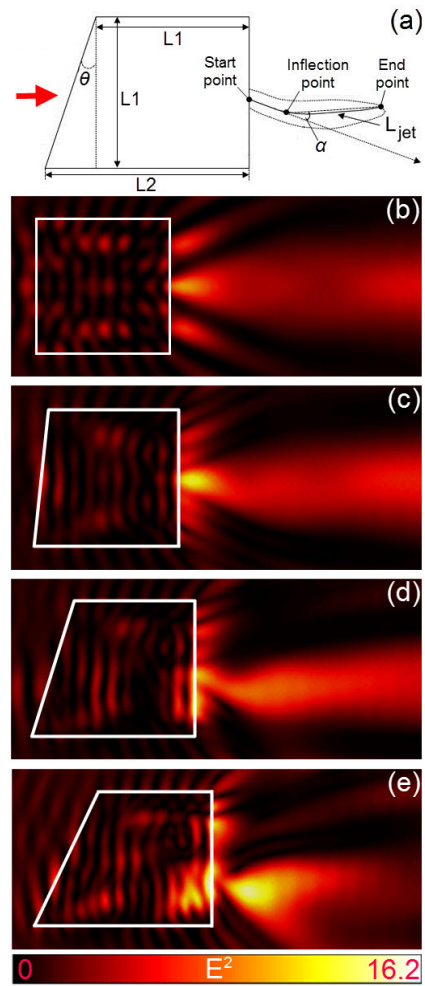

**Figure 1**. Numerical analysis of PH formation behind dielectric particle: (a) schemes of ADP (b-e) numerically-obtained spatial distributions of EM field intensity $E^2(r)$ (r is a radius vector) behind the symmetric and asymmetric dielectric cuboids, featuring equal dimensions $L1/\lambda$ and possessing the refractive index $n$ = 1.46 with changes in the particle geometry ($L1$ vs. $L2$): 4.8 mm vs. 4.8 mm, 4.8 mm vs. 5.28 mm, 4.8 mm vs. 6.4 mm, and 4.8 mm vs. 7.68 mm

The photonic hook's shape and curvature radius can be adjusted by varying wavelength,

incident light polarization and geometric and optical parameters of the emitting particle. For example, in the Fig.2, the numerical simulations of PH formation behind dielectric particle with fixed values $\theta$ and dimensions for different refractive index of particle material are shown. The PH entirely matches the trend of power flows in a certain shadow region.

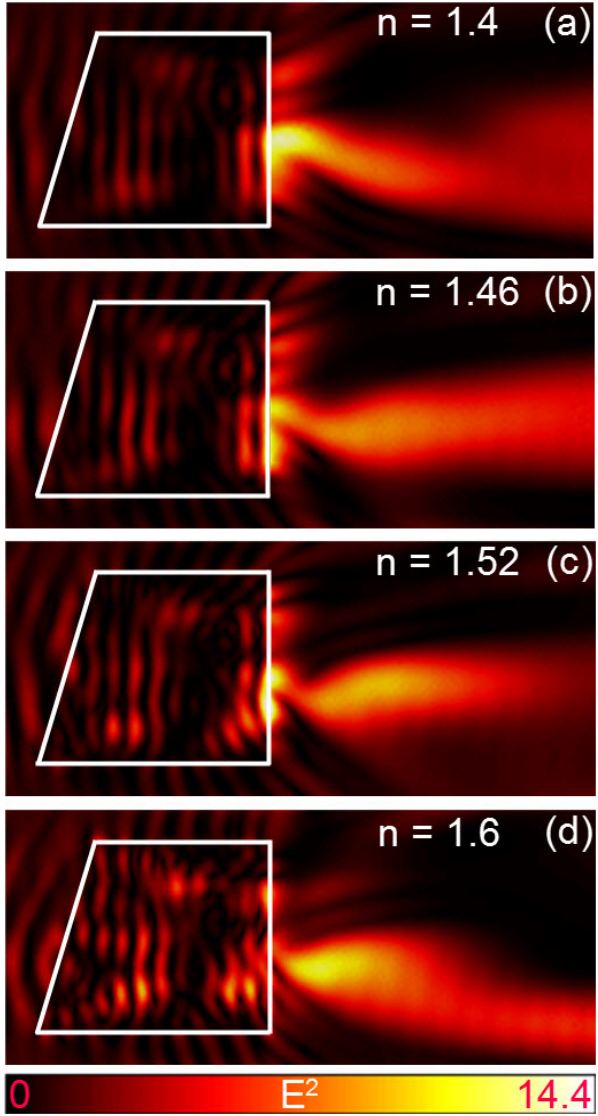

**Figure 2**. Numerical simulations of PH evalution behind ADP with optimal θ and different refractive index of particle material, which shown in inserts.

In order to experimentally showcase the PH phenomenon, we fabricated the developed dielectric particle with an optimal geometry from a bulk piece of TPX by its mechanical shaping using a milling machine without subsequent polishing of the surface. To probe the 3D structure of EM field formed behind this particle, we assembled an experimental setup relying on the principles of fiber-based scanning-probe THz imaging [16]; see Fig. S1. As a source of continuous-wave THz radiation, we use a backward-wave oscillator equipped with a wire-grid linear polarizer. As a detector we use a Golay cell. The linear polarization of the THz wave is oriented to ensure efficient generation of the PH – namely, the electric field is directed transversely to the larger facet of the dielectric particle. In order to visualize the spatial distribution of electromagnetic field intensity formed, behind the cuboid, we used an optical probe made of a flexible monocrystalline sapphire fiber [17,18] with the diameter of 300 μm, the length of 300 mm, and the flat ends. Thanks to rather low absorption of THz wave in sapphire [19], this fiber allows for guiding the THz wave over tens centimeters in a broad spectral range [20]. Furthermore, high refractive index of sapphire [19] leads to strong confinement of guiding modes in a fiber and, thus, yields sub-wavelength resolution in imaging. The lateral spatial resolution of such fiber-based imaging approach is limited by the fiber diameter (~300-μm-resolution for the employed fiber) and the depth resolution is limited by the step and accuracy of the raster-scan; thus, we could expect it to be 100 μm.

In Fig. 3, we show the results of PH visualization. It can be seen (Table 1) that curvature of PH is about $\alpha = 148^0$, the length of PH is less that 2 wavelength with inflection point position near $Z=1.2\lambda$ from the shadow surface of particle. It is important to note that the PH curvature radius is smaller than its wavelength (see Figs. 1 and 3, and Tab. 1). On the other hand the Airy beam's characteristic diameter is related to the lens numerical aperture, usually amounting to tens of wavelengths, with the Airy beam's path length being related to the optical element diameter [2,3].

PH also has a cross-dimension smaller than the wavelength that makes high resolution possible (Table 1). Thereby, the observed PH phenomenon can provide advantages over common photonic jets in imaging applications, similarly to the beneficial character of the Airy beams over the Gaussian and Bessel beams reported in Ref. [22]. Being compared to a symmetric photonic jet, PH should lead to an increased spatial resolution and an enhanced field of view. Although, such an application of PH seems to be a subject of additionally comprehensive study.

The PHs appear even in free space and require no waveguiding structures or external potentials. We believed that the curved trajectory of PH is beneficial to photonics and the related disciplines. It could be used for advanced manipulation of nanoparticles, biological systems, material processing and surgical systems, where curved sub-wavelength-scale optical caustics are of interest.

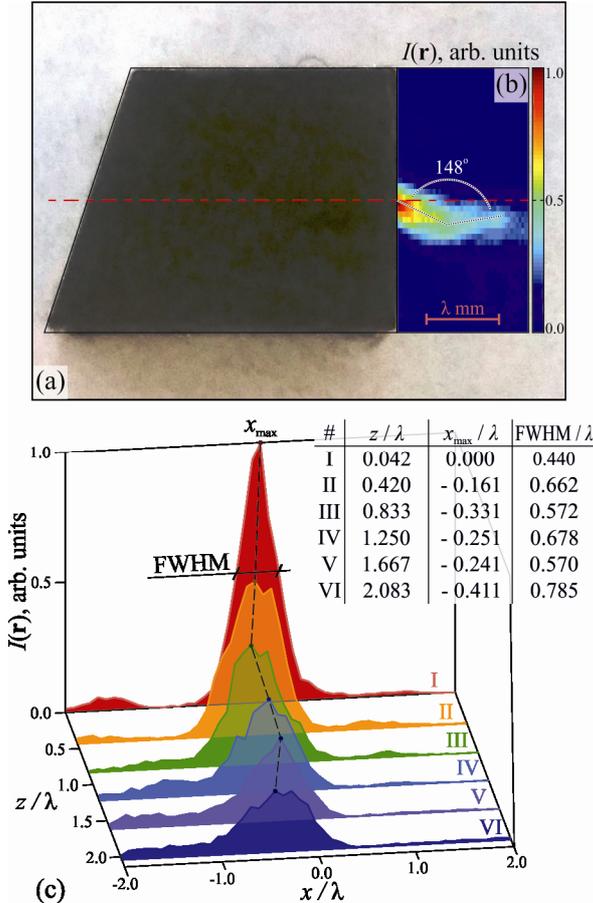

**Figure 3**. Experimental visualization of a photonic hook (with parameters from Fig.1d): (a) a photo of dielectric particle and a THz field intensity $I(\mathbf{r})$, formed at its shadow side and illustrating PH effect, (b) cross-sections of the observed PH. A table-inset in panel (c) shows parameters of the PH cross-sections

Particularly, using numerical methods, it was recently shown that the trajectory of a probe nanoparticle placed in PH is equally curved [21] and we can make a manipulator to move particles along a curved path around transparent obstacles on a nanoscale.

In contrast to traditional Airy beams, generated using a complicated optical element with cubic phase or with a spatial light modulator behind the focus of spherical lens, PH can be created using a compact mesoscale optical element (dielectric particle-lens). Moreover, Airy beam consists of a main lobe and a family of side beamlets whose intensity decay exponentially [2,3]. In the case of PH only main lobe has a curved shape and the family of curved sidelobes is absent. It was also shown that PH phenomenon is observed on a scale much smaller than Airy beams. The mesoscale dimensions and simplicity of the PH are much more controllable for practical tasks and could enable it to be integrated, for example, into lab-on-a-chip platforms and indicating their large scale potential applications. This property could be also used to redirect an optical signal in meso- and nano-scale and as a beam splitter or even as a scalpel tips for potential use in ultraprecise laser surgery or for the short-range micro manipulation.

It could be noted that the simple geometry of a particle considered in this paper is not the only possible one and serves as one of the examples of the realization of the effect under consideration. Optimization of the shape of the photon hook and its characteristics depending on the specific application is possible due to the choice of particles of a different shape.

Finally, we would like to stress that the observed PH phenomenon has a general (multidisciplinary) character and should be inherent to acoustic and surface waves, electron, neutron, proton and other beams interacting with asymmetric mesoscale obstacles. Observation of the PH phenomenon for a different types of asymmetric mesoscale dielectric particles, searchin for its application in various branches of optics and photonics, as well as in related branches of science, seem to be a prospective topics for further research.

Acknowledgements

Experimental study by K.I.Z. and N.V.Ch. was supported by Russian Science Foundation, Project # 17-79-20346, L.Y and Z.W. was supported by Sêr Cymru National Research Network (NRNF66), UK, O.V.M. was partially supported by the Mendeleev scientific fund of Tomsk State University, and I.V.M. by Tomsk Polytechnic University Competitiveness Enhancement Program grant.